\documentstyle[eqsecnum,preprint,aps]{revtex}
\newcommand{\be}{\begin{equation}}
\newcommand{\ee}{\end{equation}}
\newcommand{\bea}{\begin{eqnarray}}
\newcommand{\eea}{\end{eqnarray}}
\newcommand{\OP}{\vec{\psi}}
\newcommand{\SG}{\vec{\sigma}}
\newcommand{\M}{\vec{m}}
\newcommand{\U}{\vec{u}}
\newcommand{\F}{{\cal F}}
\newcommand{\A}{\omega_{0}}
\newcommand{\B}{\omega_{2}}
\newcommand{\C}{\omega_{1}}
\begin{document}
\draft
\title{Fluctuations and defect-defect correlations \\ in the ordering 
kinetics of the $O(2)$ model}
\author{Gene F. Mazenko and Robert A. Wickham}
\address{The James Franck Institute and the Department of Physics\\ The
University of Chicago\\ Chicago, Illinois 60637}
\date{\today}
\maketitle
%
%
\begin{abstract}
The theory of phase ordering kinetics for the $O(2)$ model using the gaussian 
auxiliary field approach is reexamined from two points of view. The effects 
of fluctuations about the ordering field are included and we organize the 
theory such that the auxiliary field correlation function is analytic in the 
short-scaled distance $(x)$ expansion. These two points are connected and we 
find in the refined theory that the divergence at the origin in the 
defect-defect correlation function $\tilde{g}(x)$ obtained in the original 
theory is removed. Modifications to the order-parameter 
autocorrelation exponent $\lambda$ are computed.
\end{abstract}
\draft
%
%
\section{Introduction}

The phase ordering kinetics 
of systems with continuous symmetry, such as the $O(n)$ model,
is particularly interesting because of the topological defect structures
produced: vortices and strings for $n=2$ and monopoles for $n=3$ 
\cite{MERMIN79}. 
While much is understood about the theory of growth kinetics 
for the $O(n)$ model, there are some interesting unresolved problems 
associated with the short-distance behaviour of the defect-defect correlation 
function $\tilde{g}(x)$. In the theory developed in \cite{MAZENKO90} one 
expresses the order parameter $\OP({\bf R},t)$ as a local non-linear function 
of an auxiliary field
$\vec{m}({\bf R},t)$ which is physically interpreted as the distance, at time 
$t$, from position ${\bf R}$ to the closest defect. 
One of the physical motivations for introducing $\vec{m}({\bf R},t)$
is that it is {\it smoother} than the order-parameter field.  Sharp 
interfaces or well defined defects produce a non-analytic structure in 
the order-parameter scaling function $\F(x)$
at small-scaled distances $x$ which, physically,  is responsible for
the Porod's law decay seen scattering experiments \cite{POROD}. The 
expectation, however, is that the auxiliary field correlation
function $f(x)$ will be analytic in this same distance range.  In the case of
a scalar order-parameter these expectations are supported by the
theory \cite{MAZENKO90}. However for $n > 1$, as pointed out in \cite{LIU92a},
this is not the case.  One finds a weak  non-analytic component in
$f$ and, more significantly, for $n=2$ one can trace this non-analytic 
component to an unphysical divergence in $\tilde{g}(x)$ at small $x$.
\cite{LIU92b}. This divergence is not seen in simulations \cite{MONDELLO90}
or experiments \cite{NAGAYA95} for $n=d=2$ where $\tilde{g}(x)$ apparently 
approaches zero at the origin.

In this paper we focus on the case $n=2$ and
show how these problems can be resolved by taking seriously the
assumption that the correlations of the auxiliary field 
are indeed smoother than those of the order parameter.  
We find that it is possible to rearrange the theory such that
$f$ is analytic in $x$ if we extend the theory to include fluctuations 
about the ordering field and treat the separation between the ordering
field and the fluctuation fields carefully. This is accomplished by 
introducing a new field $\vec{\Theta}$ which is constructed to ensure that 
the fluctuations are small, while at the same time compensating for the 
non-analyticities in $f$. It is important to note that we work at 
zero-temperature, so the fluctuations are not thermally driven \cite{KRAMER94}.
Rather, we will see that the correlations in the fluctuations are slaved to the
correlations in the order parameter. The theory of Ohta, Jasnow and
Kawasaki (OJK) \cite{OJK} basically avoids this entire discussion by simply 
assuming that $f$ has a gaussian form \cite{CONFUSION}. There is no 
self-consistent determination of $f$ in that theory.

It is well
established that for late times following a quench from the disordered to the
ordered phase the dynamics obey scaling and the system can be described in 
terms of a single growing length $L(t)$, which is characteristic of the 
spacing between defects. In this scaling regime the order-parameter 
correlation function
%
%
\be
\label{EQ:OPCOR}
C(12) \equiv \langle \OP (1) \cdot \OP (2) \rangle
\ee
has a universal equal-time scaling form
%
%
\be 
C(12) = \psi_{0}^{2} \F(x), 
\ee
where $\psi_{0}$ is the magnitude $\psi = |\OP|$ of the order-parameter in the
ordered phase.  Here we use the short-hand notation where $1$ denotes  
$({\bf R_{1}},t_{1})$, and define the scaled length $x$ as 
$x = R/L(t)$ with $R \equiv |{\bf R}| \equiv |{\bf R_{2}} - {\bf R_{1}}|$. 
It is also well established that, in the scaling regime, 
 $L(t) \sim t^{\phi}$ where $t$ is the time after the quench.
For the non-conserved models considered here the exponent $\phi = 1/2$. 
Another measurable quantity is the exponent $\lambda$ governing 
the decay of order-parameter autocorrelations, and defined by
%
%
\be
\label{EQ:LAMDEF}
C({\bf 0},t,t') \sim \frac{1}{L^{\lambda}(t)} \mbox{ for } t \gg t'.
\ee
This non-trivial exponent can be computed theoretically, along with the 
scaling function $\F$ \cite{LIU92a}. The predictions for $\lambda$ are in 
excellent agreement with simulation results \cite{LIU91,LEE95}. 
The theoretical predictions for $\F$ are also in good
agreement with simulations \cite{MAZENKO90}.

The dynamics of the defect structures themselves is amenable to 
theoretical treatment \cite{LIU92b}. In this paper we shall mainly be 
interested in the 
case $n=d$, where the defects are points. For $n=d-1$ the defects are 
strings, but the analysis follows closely that for 
point defects and yields qualitatively similar results. The density of 
point defects, for $n=d$, is defined as
%
%
\be
\rho(1) = \sum_{\alpha} q_{\alpha} 
          \delta({\bf R_{1}} - {\bf x_{\alpha}}(t_{1}))
\ee
where ${\bf x_{\alpha}}(t_{1})$ is the position at time $t_{1}$ of the 
$\alpha$th point defect, which has a charge $q_{\alpha}$. 
Correlations in $\rho$,
%
%
\be
G(12) \equiv \langle \rho (1) \rho(2) \rangle,
\ee
at equal-times $t_{1}=t_{2}=t$ can be shown \cite{LIU92b} to decompose into 
two parts
%
%
\be
\label{EQ:POINTDEF}
G({\bf R}, t) = n_{0}(t) \delta({\bf R}) + g({\bf R},t).
\ee
The first term $n_{0}(t)$ represents defect self-correlations and is just
the total unsigned number density of defects at time $t$. We will be 
primarily concerned here with the second term $g({\bf R},t)$ which measures 
the correlations between different defects. In the scaling regime
it can be shown \cite{LIU92b} that $n_{0}(t) \sim L^{-n}(t)$ and that 
$g({\bf R},t)$ has the form
%
%
\be
\label{EQ:DEFDEF}
g({\bf R},t) = \frac{1}{L^{2n}(t)} \tilde{g}(x).
\ee
where $\tilde{g}(x)$ is a universal scaling function.

In the next section we present the $O(n)$ model and describe the mapping 
between the auxiliary field and the order parameter. In Section 
III we discuss the separation of the equation of motion into an equation 
for the evolution of the ordering field and an equation for the 
dynamics of the fluctuations. The main analytical 
results of the paper are presented in Section IV, where we discuss how 
the quantities $\F$, $\lambda$ and $\tilde{g}$ are determined through the 
solution of a non-linear eigenvalue problem. In Section V we calculate the
correlations in the fluctuations, assuming that the 
fluctuation field $\U$ and the auxiliary field $\M$ form a set of coupled
gaussian variables.
Our numerical analysis of the new non-linear eigenvalue problem 
is presented in 
Section VI and the results are discussed in the final section, which also
addresses more general issues that indicate directions for future 
research. 
%
%
\section{Model}

We consider the $O(n)$ model, which describes the dynamics
of a  non-conserved, $n$-component order-parameter field $\OP (1) =  
(\psi_{1} (1), \cdots,\psi_{n} (1) )$. To begin we will work with general $n$;
however, later we will focus on the interesting case $n=2$. 
As in previous work in this area \cite{LIU92a}, the dynamics are 
modeled using a time-dependent Ginzburg-Landau equation
%
\be
\frac{\partial \OP}{\partial t} = - \Gamma \frac{\delta F[\OP]}
{\delta \OP}.
\label{EQ:LANG}
\ee
We assume that the quench is to zero temperature where the usual noise 
term on the right-hand side is zero \cite{BRAY89}.
 $\Gamma$ is a kinetic coefficient and 
$F[\OP]$ is the free-energy, assumed to be of the form
%
\be
F[\OP] = \int d^{d} r ( \frac{c}{2} |\nabla \OP|^{2} + V[\psi])
\ee
where the potential $V[\psi]$ is chosen to have $O(n)$ symmetry and a 
degenerate ground state with $\psi = \psi_{0}$. Since only these properties 
of $V$ will be important in what follows we need not be more specific in our 
choice for $V$ \cite{POTENTIAL}. With a suitable redefinition of the time and 
space scales the coefficients $\Gamma$ and $c$ can be set to one and 
(\ref{EQ:LANG}) can be written as 
%
\be
\frac{\partial \OP}{\partial t} = \nabla^{2} \OP - 
\frac{\partial V[\psi]}{\partial \OP}.
\label{EQ:MOT}
\ee
It is believed that our final results are independent of the exact nature of
the initial state, provided it is a disordered state.

The evolution induced by
(\ref{EQ:MOT}) causes $\OP$ to order and assume a distribution that is far 
from gaussian. It is by now standard to introduce a mapping between the 
physical field $\OP$ and an auxiliary field $\M$ with more 
tractable statistics. We can decompose $\OP$ exactly as
%
\be
\label{EQ:MAPPING}
\OP = \SG [ \M ] + \U. 
\ee
The utility of this decomposition lies in our ability to create a 
consistent theory with the mapping $\SG$ chosen to reflect the defect
structure in the problem, and $\U$ constructed to be small at late-times.
Thus $\U$ represents fluctuations about the ordering field $\SG$.   
The precise statistics of the fields $\M$ and $\U$  will be specified below. 

The defect structure \cite{DEFECT} is naturally incorporated by using the 
Euler-Lagrange equation for the order-parameter around a static defect in 
equilibrium,
%
\be
\nabla^{2}_{m} \SG [\M] = \frac{\partial V [\SG]}{\partial \SG },
\label{EQ:SIGPOT}
\ee
to determine the functional dependence of $\SG$ on $\M$.
The defects are then the non-uniform solutions of (\ref{EQ:SIGPOT}) which 
match on to the uniform solution at infinity. Since we expect only the 
lowest-energy defects, having unit topological charge, will survive to
late-times the relevant solutions to (\ref{EQ:SIGPOT}) will be of the form
%
\be 
\SG [\M] = A(m) \hat{m}
\label{EQ:SIGEX}
\ee
where $m = |\M| \mbox{ and } \hat{m} = \M/m$. Thus the interpretation of 
$\M$ is that its magnitude represents the distance away from a defect core and
its orientation indicates the direction to the defect core. We expect 
$m$, away from the defect cores, to grow as $L$ in the 
late-time scaling regime. Inserting (\ref{EQ:SIGEX}) into (\ref{EQ:SIGPOT})
gives an equation for $A$,
%
\be
\nabla^{2}_{m} A - \frac{n-1}{m^{2}} A - V'[A] = 0,
\label{EQ:A}
\ee
where the prime indicates a derivative with respect to $A$.
The boundary conditions are $A(0) = 0, \mbox{ } A(\infty) = \psi_{0}$. An 
analysis of (\ref{EQ:A}) for $n>1$ and large $m$ yields
%
\be
A(m) = \psi_{0} \left[ 1 - \frac{\kappa}{m^{2}} + \cdots \right]
\label{EQ:A-LGM}
\ee
where $\kappa = (n-1)/V''[\psi_{0}] > 0$.
The algebraic relaxation of the order-parameter to its ordered value is a
distinct feature of the $O(n)$ model for $n>1$. In the scalar case $(n=1)$
$\psi$ relaxes exponentially to $\psi_{0}$ away from the defects.

We shall also be interested in the stability matrix defined by
%
%
\bea
\label{EQ:STABILITY}
W_{ij}[\SG ] & \equiv & \frac{\partial ^{2}V[\SG ]}{\partial \sigma_{i}
\partial \sigma_{j}} \\
\nonumber
& = & V''[A]\hat{\sigma}_{i}\hat{\sigma}_{j} +\frac{V'[A]}{A}
(\delta_{ij}-\hat{\sigma}_{i}\hat{\sigma}_{j}).
\eea
By the definition
of $\psi_{0}$ and what we mean by equilibrium 
we have
\be
V'[\psi_{0}] =0
\ee
and
\be 
V''[\psi_{0}]  \equiv q_{0}^{2} > 0.
\ee
These results give
\be
W_{ij}[\psi_{0} ] =q_{0}^{2}\hat{\sigma}_{i}\hat{\sigma}_{j}
\ee
which is purely longitudinal. This reflects the fact that, in
equilibrium, the longitudinal fluctuations have a ``mass'' $q_{0}^{2}$ while 
the transverse fluctuations, or spin-waves, are massless.
%
%
\section{Separation of Equations of Motion}

In this section we develop the equations of motion satisfied by
the fields $\SG$ and $\U$.  Let us first define
%
%
\be
H_{i}[\OP]\equiv \frac{\partial V[\OP]}{\partial \psi_{i}}
\ee
and rewrite the equation of motion (\ref{EQ:MOT}) in the form
%
%
\be
\label{EQ:EXPLICITSEP}
\frac{\partial}{\partial t}(\SG + \U)= 
\nabla^{2}(\SG + \U) -\vec{H}[\SG + \U ].
\ee
We then quite generally assume that $\SG$ satisfies the equation of
motion
%
%
\be
\label{EQ:SIGMAONLY}
\frac{\partial \SG}{\partial t} = \nabla^{2}\SG 
-\nabla_{m}^{2}\SG + \vec{\Theta} 
\ee
where $\vec{\Theta}$ is as yet unspecified.  Clearly, subtracting this
equation from the equation of motion  (\ref{EQ:EXPLICITSEP}) and using 
(\ref{EQ:SIGPOT}) we obtain
%
%
\be
\label{EQ:NLU}
\frac{\partial \U}{\partial t} = \nabla^{2} \U 
-\vec{H}[\SG + \U ] + \vec{H}[\SG ] - \vec{\Theta}. 
\ee
To this point things are quite general since we have not specified
$\vec{\Theta}$.  A key point is that $\vec{\Theta}$ must be chosen
such that $\U$ does indeed represent a fluctuation.  This means that
in the scaling regime we can treat $\U$ as small and keep only
leading powers of $\U$ in the equations of motion for $\SG$ and
$\U$.  Equation (\ref{EQ:NLU}), to leading order, is then given by
%
%
\be
\label{EQ:UONLY}
\frac{\partial u_{i}}{\partial t} = \nabla^{2}u_{i} 
-W_{ij}[\SG] u_{j} - \Theta_{i}  
\ee
where a sum over the index $j$ is assumed.

We now assume that $\vec{\Theta}$ is a function of $\M$ only.  This 
means that $\SG$ satisfies a closed equation, while $\U$
is slaved by $\M$.  We will choose the form for $\vec{\Theta}$
so that the correlation function $f$ for $\M$ is analytic for short-scaled 
distances.  As we shall see this is a rather constrained
process.
%
%
\section{Analysis of the $\SG$ degrees of freedom}

\subsection{Construction of $\vec{\Theta}$}

If we set $\vec{\Theta}$ equal to zero in (\ref{EQ:SIGMAONLY}) we obtain the 
equation used previously to determine the $\SG$ correlations \cite{LIU92a}.
This choice decouples $\SG$ and $\U$. The equation for $\U$ would then
separate into a (massless \cite{MASS}) diffusion equation for the transverse 
piece $\U_{T}$ and
an equation for the longitudinal piece $u_{L}$ with a mass term 
$-q_{0}^{2} u_{L}$.
However, the equation for $\SG$ would necessarily lead to non-analytic 
behaviour in $f$ at short-scaled distances and ultimately to an unphysical
divergence in $\tilde{g}(x)$ at small $x$. We must
choose $\vec{\Theta}$ so that $f(x)$ is analytic for small $x$.  
The form we can use for $\vec{\Theta}$ is determined by the following 
observations:

(i)  $\vec{\Theta}$ must be odd under $\M \rightarrow -\M$.

(ii)  $\vec{\Theta}$ must scale as $ {\cal O}(L^{-2})$ in the
scaling regime if it is to compensate for the terms in the
equation of motion which lead to the non-analyticities in 
$f$. This will also allow us to treat $\vec{u}$ as a fluctuation 
since it will imply $\vec{u} \sim L^{-2}$.

It is not easy to construct a variety of functions of $\M$ which are 
independent and
satisfy (i)
and (ii).  We propose the general form
%
%
\be
\label{EQ:THETADEF}
\vec{\Theta} =\frac{\A}{L^{2}(t)} \SG  
+\sum_{\ell =1}^{\ell_{max}} \omega_{\ell} [\nabla \M]^{2(\ell-1)} 
\nabla^{2}_{m} \SG
\ee
where $[\nabla \M]^{2} = \sum_{i=1}^{d} \sum_{\alpha = 1}^{n} 
[\partial_{i} m_{\alpha}]$ and all of the $\omega_{\ell}$, $\ell \geq 0$ are 
assumed to be of $ {\cal O}(1)$. One can think of including 
other quantities like $(\psi_{0}^{2}-\SG^{2})\SG$ but these, in the 
scaling regime, are equivalent to $\nabla_{m}^{2} \SG$.
It is interesting to note, using the definition (\ref{EQ:SIGPOT}) for
$\nabla^{2}_{m}\SG$,  that $\vec{\Theta}$ is longitudinal.

For the purposes of this paper we will only consider constructing $f(x)$
to be analytic through terms of ${\cal O} (x^{4})$. To satisfy this 
requirement it is sufficient to set $\ell_{max} = 2$ in (\ref{EQ:THETADEF}). 
The equation for $\SG$ (\ref{EQ:SIGMAONLY}) is then of the form 
%
%
\be
\label{EQ:SIGMAMOT}
\vec{B} = 0
\ee
where we define, for later convenience,
%
%
\be
\vec{B} \equiv \partial_{t} \SG  - \nabla^{2} \SG 
+ \nabla_{m}^{2} \SG - \frac{\A}{L^{2}(t)} \SG - \C 
\nabla_{m}^{2} \SG - \B [\nabla \M]^{2} \nabla_{m}^{2} \SG.
\ee
%
%
\subsection{The Gaussian Approximation}

To complete the definition of the model one must specify the form of 
the probability distribution for the auxiliary field $\M$. Forcing
$\SG$ to satisfy the exact equation of motion (\ref{EQ:SIGMAMOT})
is tantamount to solving the problem exactly, and will determine a 
probability distribution for $\M$ which is complicated and extremely 
difficult for purposes of computation. Progress can be made if one 
imposes the weaker constraint
%
\be
\label{EQ:TWOPTCONST}
\langle \vec{B}(1) \cdot \SG (2) \rangle = 0.
\ee
This equation allows one to insure that $\vec{B}(1)$ is reasonably small at 
late-times but gives one the flexibility to choose a suitable probability
distribution. The simplest choice is a gaussian probability distribution for
$\M$ with the correlation function $C_{0} (12)$  explicitly defined through
%
\be
\langle m_{i} (1) m_{j} (2) \rangle = \delta_{ij} \mbox{ } C_{0} (12).
\ee
The system is assumed to be statistically isotropic and homogeneous so 
$C_{0}(12)$ is invariant under interchange of its spatial indices. For future
reference we also define the one-point correlation function
%
%
\be
S_{0}(1) = C_{0}(11)
\ee
and the normalized correlation function
%
%
\be
f(12) = \frac{C_{0}(12)}{\bar{S}_{0}(12)}
\ee
with $\bar{S}_{0}(12) = \sqrt{S_{0}(1)S_{0}(2)}$. As discussed above it is 
expected that both $C_{0}$ and $S_{0}$ grow as $L^2$ at late times. 
The gaussian approximation, 
which has been successful in describing the correlations in 
these systems, forms the basis of 
almost all present analytical treatments of phase-ordering problems
\cite{MAZENKO90,LIU92a}. Efforts to go beyond to gaussian approximation
are defined in \cite{MAZENKO94,WICKHAM}.

The functional dependence of $C(12)$ and $\tilde{g}(x)$ on $f$ 
can be derived without reference to the dynamics contained in 
(\ref{EQ:TWOPTCONST}).
Using (\ref{EQ:MAPPING}). (\ref{EQ:SIGEX}), and (\ref{EQ:A-LGM}) 
$C(12)$ can be written to leading order in $1/L$:
%
%
\bea
\label{EQ:OPCOR2}
C(12) & = & \psi_{0}^{2} \langle \hat{m} (1) \cdot \hat{m} (2) \rangle \\
\nonumber     & = & \psi_{0}^{2} \F(12).
\eea
Assuming gaussian statistics for $\M$ we obtain \cite{LIU92a,BRAY91}
%
%
\be
\label{EQ:HYPER}
\F(12) = \frac{nf(12)}{2 \pi} B^{2} \left[ \frac{1}{2},\frac{n+1}{2} \right]
F \left[ \frac{1}{2}, \frac{1}{2};\frac{n+2}{2};f^{2} (12)\right]
\ee
where $B$ is the beta function and $F$ is the hypergeometric function.
Within the gaussian theory, $\tilde{g}(x)$ is given by \cite{LIU92b} 
%
%
\be
\label{EQ:DEFECTDEFECTCOR}
\tilde{g} (x) = n! \left[ \frac{h}{x} \right]^{n-1} 
\frac{\partial h}{\partial x}
\ee
with $h = - \gamma f'/2 \pi$ and $\gamma = 1/\sqrt{1-f^2}$. The defect 
density is given by
%
%
\be
n_{0}(t) = \frac{n!}{2^{n} \pi^{n/2} \Gamma(1 + n/2)} \left[ 
\frac{S_{0}^{(2)}}{n S_{0}(t)} \right]^{n/2}
\ee
with 
%
%
\be
\label{EQ:DEFS02}
S_{0}^{(2)} = \frac{1}{n} \langle [\nabla \M ]^{2} \rangle.
\ee
%
%
%
%
%
%
\subsection{Order Parameter Correlations}

With the specification of the probability distribution for $\M$ the 
constraint (\ref{EQ:TWOPTCONST}) leads to an equation that allows one to 
compute correlations in the order parameter $\SG$. The quantity 
$S_{0}^{(2)}$ (\ref{EQ:DEFS02}), which will later appear 
in the definition of the length scale and in the formula for the 
autocorrelation exponent $\lambda$, is determined through condition 
(\ref{EQ:TWOPTCONST}), with $2=1$:
%
%
\bea
\nonumber
\frac{1}{2} \partial_{t_{1}} \langle \SG^{2} (1) \rangle - \langle \SG (1) 
\cdot \nabla_{1}^{2} \SG (1) \rangle - \frac{\A}{L^{2} (t_{1}) } \langle
\SG^{2} (1) \rangle  & + (1 - \C) \langle \SG (1) \cdot \nabla_{m}^{2} 
\SG (1) \rangle & \\ 
& - \B \langle \SG (1) \cdot [\nabla_{1} \M(1)]^{2} \nabla_{m}^{2} 
\SG (1) \rangle &  = 0.
\label{EQ:ONEPTEOM}
\eea
Equation (\ref{EQ:ONEPTEOM}) can be simplified by using the following 
identities, true for gaussian averages,
%
%
\bea
\langle \SG(1) \cdot \nabla_{1}^{2} \SG (1) \rangle & = &
- S_{0}^{(2)} \langle \nabla_{m} \cdot [ \SG (1) \cdot \nabla_{m} \SG (1) ]
\rangle + S_{0}^{(2)} \langle \SG (1) \cdot \nabla_{m}^{2} \SG (1) \rangle
\\
\langle \SG(1) \cdot [ \nabla_{1} \M (1) ]^{2} \nabla_{m}^{2} \SG(1) \rangle
& = & n S_{0}^{(2)} \langle \SG (1) \cdot \nabla_{m}^{2} \SG (1) \rangle,
\eea
and observing that, at late-times, the dominant term in (\ref{EQ:ONEPTEOM}) is
%
%
\bea
\nonumber
\langle \SG (1) \cdot \nabla_{m}^{2} \SG (1) \rangle & = &
- \frac{\psi_{0}^{2}}{2 S_{0} (1) } \ln{S_{0}(1)} \mbox{ for } n = 2 \\
& = & - \frac{\psi_{0}^{2}}{S_{0}(1)} \frac{n-1}{n-2} \mbox{ for } n >  2.
\eea
Evaluating (\ref{EQ:ONEPTEOM}) to leading order in $1/L$ one has
%
%
\be
\label{EQ:S02DEF}
S_{0}^{(2)} = \frac{1 - \C}{1 - 2 \mu (n - 2) \A  / \pi (n-1) + n \B}
\ee
where we have defined the scaling length
%
%
\be
\label{EQ:LENGTH}
L^{2}(t) = \frac{\pi S_{0}(t)}{2 \mu S_{0}^{(2)}} = 4 t.
\ee
Note that for $n=2$ the term with $\A$ does not appear in $S_{0}^{(2)}$
because it is dominated by  the ${\cal O}(L^{-2} \ln L)$ terms 
in (\ref{EQ:ONEPTEOM}). There is a further simplification of 
(\ref{EQ:S02DEF}) for $n=2$ since later we will have to set
$\C + 2 S_{0}^{(2)} \B = 0$ to ensure that the correlations in $\U$ 
remain finite. With this relation we have
%
%
\be
\label{EQ:SIMSO2}
S_{0}^{(2)} = 1 \mbox{ for } n=2,
\ee
which is the value for $S_{0}^{(2)}$ obtained previously for all $n$ when
$\vec{\Theta}=0$.

Equation (\ref{EQ:TWOPTCONST})  directly determines the time evolution of the 
two-point order-parameter correlations and is given explicitly by:
%
%
\bea
\nonumber
\left[ \partial_{t_{1}} - \nabla_{1}^{2} - \frac{\A}{L^{2}(t_{1})} 
\right] \langle \SG (1) \cdot \SG (2) \rangle & + (1 - \C) \langle 
\nabla_{m}^{2} \SG (1) \cdot \SG (2) \rangle  & \\
 & - \B \langle \nabla_{m}^{2} \SG (1) [\nabla_{1} \M (1) ]^{2} 
\cdot \SG (2) \rangle & = 0.
\label{EQ:AVGEOM}
\eea
Equation (\ref{EQ:AVGEOM}) can be re-expressed as an equation for  
the order-parameter correlation function $\F$ (\ref{EQ:HYPER}) 
by means of the following identities:
%
%
\bea
\langle \nabla_{m}^{2} \SG (1) [\nabla_{1} \M (1) ]^{2} \cdot \SG (2) \rangle
& =  & n S_{0}^{(2)} \langle \nabla_{m}^{2} \SG (1) \cdot \SG (2) \rangle
+ [ \nabla_{1} C_{0} (12) ]^{2} \langle \nabla_{m}^{2} \SG (1) \cdot 
\nabla_{m}^{2} \SG (2) \rangle \\
\langle \nabla_{m}^{2} \SG (1) \cdot \SG (2) \rangle & = & 
\frac{-\psi_{0}^{2}}{S_{0}(1)} f \partial_{f} \F \\
\langle \nabla_{m}^{2} \SG (1) \cdot \nabla_{m}^{2} \SG (2) \rangle & = &
\frac{\psi_{0}^{2}}{\bar{S}_{0}^{2}} ( f \partial_{f} \F + 
f^{2} \partial_{f}^{2} \F ),
\eea
where we use the shorthand notation $f = f(12)$, $\F = \F(12)$, and 
$\partial_{f} \F = \partial \F / \partial f$ {\em etc}.
Equation  (\ref{EQ:AVGEOM}) becomes
%
%
\be
\label{EQ:AVGEOM2}
\left[ \partial_{t_{1}} - \nabla_{1}^{2} - \frac{\A}{L^{2}(t_{1})} \right] 
\F - \frac{1 - \C - n S_{0}^{(2)} \B}{S_{0}(1)} f \partial_{f} \F - 
\B [ \nabla_{1} f ]^{2} [f \partial_{f} \F + f^{2} \partial^{2}_{f} \F] = 0,
\ee
which is the starting point for the evaluation of the two quantities of 
interest here: the autocorrelation exponent $\lambda$ (\ref{EQ:LAMDEF})
and the late-time scaling form for $\F$.

For times $t_{1} \gg t_{2}$ both $\F$ and $f$ are small. In 
this limit (\ref{EQ:AVGEOM2}) becomes a linear equation for $\F$ and,
following the treatment in \cite{LIU92a,LIU91}, with the definition 
(\ref{EQ:LENGTH}), $\lambda$ can be determined as:
%
%
\be
\label{EQ:LAMBDA}
\lambda = d - \frac{\pi}{4 \mu} 
\frac{1 - \C - n S_{0}^{(2)} \B}{S_{0}^{(2)}} - \frac{\A}{2}.
\ee
If one knows $\A, \mbox{ } \C,\mbox{ } \B$ and $\mu$ one can determine 
$\lambda$. These
quantities can be found from an analysis of the equal-time correlations, to 
which we now turn.

To examine the equal-time order-parameter correlations in the late-time 
scaling regime we set $t_{1}=t_{2}=t$ and write (\ref{EQ:AVGEOM2}) in terms
of the scaled distance $x$. To leading order in $1/L$ we have 
%
%
\be
\label{EQ:SCALE}
\vec{x} \cdot \nabla_{x} \F + \nabla_{x}^{2} \F + \A \F + 
\frac{\pi}{2 \mu} \frac{1 - \C - n S_{0}^{(2)} \B}{S_{0}^{(2)}}
f \partial_{f} \F + \B [\nabla_{x} f]^{2} [ f \partial_{f} \F + 
f^{2} \partial_{f}^{2} \F ] = 0.
\ee
The calculation of the scaling form for $\F$  reduces to the solution of
the non-linear eigenvalue problem (\ref{EQ:SCALE}) with the eigenvalue $\mu$.
The eigenvalue is selected by finding numerically the solution of 
(\ref{EQ:SCALE}) which satisfies the analytically determined boundary behaviour
at both large and small $x$. The new aspect to the problem is the presence 
of the unknowns $\A$, $\C$ and $\B$, a consequence of the incorporation of 
fluctuations into the model. For $n=2$ these constants play the role of 
counter-terms that cancel out the small-$x$ non-analyticities in the 
normalized 
auxiliary field correlation function $f$. This procedure fixes 
$\A$, $\C$ and $\B$ in terms of $\mu$ and $d$.
  
For large $x$ both $\F$ and $f$ are small and 
(\ref{EQ:SCALE}) can be linearized. In this regime the solution to 
(\ref{EQ:SCALE}) is
%
%
\be
\label{EQ:LARGEX}
\F \sim x^{d - 2 \lambda} e^{-x^{2}/2}.
\ee
The result for the exponent $d - 2 \lambda$ appears to be robust.
Until now, we have derived results valid for arbitrary $n > 1$. However, 
the primary goal of this paper is to examine the $O(2)$ model, where there
are known qualitative discrepancies with simulation data. With this in 
mind, we now  examine the small-$x$ behaviour of the scaling equation 
(\ref{EQ:SCALE}) for the case $n=2$. For small-$x$ (\ref{EQ:SCALE})
admits the following general expansion for $f$: 
%
%
\be
\label{EQ:EXPf}
f = 1 + f_{2} x^{2} \left[ 1 + \frac{K_{2}}{\ln x} \left( 1 + {\cal O} 
\left[\frac{1}{\ln x} \right] \right) \right] + f_{4} x^{4} \left[
1 + \frac{K_{4}}{\ln x} \left( 1 + {\cal O} \left[\frac{1}{\ln x} \right] 
\right) \right] + {\cal O} (x^{6}).
\ee
Non-analyticities appear as a result of the non-zero $K_{2}$ and $K_{4}$ 
coefficients multiplying factors of $1/\ln x$. The non-zero 
$K_{2}$ coefficient is particularly important since it is responsible for 
the divergence of the defect-defect correlation function at small $x$. 

The coefficients of the expansion (\ref{EQ:EXPf}) can be 
determined by examining 
(\ref{EQ:SCALE}) order-by-order at small $x$. Balancing terms at 
${\cal O} (\ln x)$ gives
%
%
\be
\label{EQ:F2DEF}
f_{2} =  - \frac{\pi}{4 \mu d}.
\ee
This relation between $f_{2}$ and $\mu$ is the same one that was found in 
the original theory \cite{LIU92a}. This equivalence is a consequence of the 
simplifications mentioned previously (\ref{EQ:SIMSO2}) that occur for $n=2$.  
At ${\cal O} (1)$ we have a equation relating $\A$, $\B$, and $K_{2}$
%
%
\be
\label{EQ:C0C2K2}
\A = 2 f_{2} ( 1 + d K_{2} + \B).
\ee
As discussed above, the constant $\C = -2 \mbox{ } \B$ for $n = 2$.
If we work with $\A=\C=\B=0$ then (\ref{EQ:C0C2K2}) implies $K_{2} = -1/d$. 
This is simply the gaussian model examined previously \cite{LIU92a}, whose
non-zero value for $K_{2}$ results in the divergent small-$x$ behaviour of 
the defect-defect correlation function $\tilde{g}(x)$. 

Now, however, we can insist that, at ${\cal O}(x^{2})$, $f$ is analytic and 
enforce $K_{2} = 0$. This choice produces the following relation between 
$\A$ and $\B$:
%
%
\be
\label{EQ:COK=0}
\A = 2 f_{2} ( 1 + \B).
\ee
The small-$x$ divergence in the defect-defect correlation function 
is now eliminated. The leading correction to the ${\cal O}(x^{2})$ term 
in $f$ is then the ${\cal O}(x^{4})$ term with a coefficient
%
%
\be
f_{4} = \frac{d + 3 + \B}{2 (d+2)} (f_{2})^{2} - 
        \frac{f_{2}(2 + \A)}{4 (d + 2)}
\ee
We can go further and insist that $f$ is analytic at small-$x$ 
up to ${\cal O}(x^{4})$. Enforcing $K_{4} = 0$ allows one to arrive at 
a complicated expression for $\B$ in terms of $\mu$ and $d$ only.

In section VI we will consider enforcing $K_{2} = 0$ through various choices 
for the $\omega_{\ell}$ and we will numerically solve the associated 
eigenvalue problem. Before doing this though we complete our discussion 
of the theory 
by examining in detail the correlations of the fluctuations and their 
relationship to the order-parameter correlations. In the process, we will
establish the important constraint on the $\omega_{\ell}$ parameters discussed
earlier.
%
%
\section{Analysis of Fluctuation Correlations}

In addition to order parameter correlations the theory also completely
describes correlations in the fluctuation field $\U$. There are two types 
of equal-time fluctuation correlations that are of interest to us. The first 
describes cross-correlations between the $\SG$ and $\U$ fields and is defined
as 
%
%
\be
\label{EQ:F1}
C_{u0}(12)= \langle \U ({\bf R_{1}},t)\cdot\SG ({\bf R_{2}},t)\rangle.
\ee
The second describes correlations of the fluctuation field with itself and is
given by
\be
\label{EQ:F2}
\delta_{ij} C_{uu}(12)= 
\langle u_{i} ({\bf R_{1}},t) u_{j} ({\bf R_{2}},t)\rangle.
\ee
As we will see later these quantities are closely related in the scaling 
regime. One can deduce equations of motion for both $C_{u0}$ and $C_{uu}$
by using the equations of motion (\ref{EQ:SIGMAONLY}) and (\ref{EQ:UONLY}) for
$\SG$ and $\U$. For equal-times one has
%
%
\bea
\nonumber
\frac{\partial}{\partial t}C_{u0}(12) & = & \nabla^{2}_{1}C_{u0}(12) -
\langle W_{ij}(1)u_{j}(1)\sigma_{i}(2)\rangle - C_{\Theta 0}(12) \\
& & +\nabla^{2}_{2}C_{u0}(12) -C_{u2}(12)+C_{u \Theta }(12)
\label{EQ:FLUCTCOR1}
\eea
and
\be
\label{EQ:FLUCTCOR2} 
\frac{1}{2} \frac{\partial}{\partial t}C_{uu}(12)= \nabla^{2}_{1}C_{uu}(12) 
- \frac{1}{n} \langle W_{ij}(1)u_{j}(1)u_{i}(2)\rangle
- \frac{1}{n} C_{\Theta u}(12)
\ee
where in the last equation we have used the translation invariance in
space to combine two equivalent terms. We have also defined
%
%
\bea
C_{\Theta 0}(12) & = & \langle \vec{\Theta} (1) \cdot \SG (2) \rangle \\
C_{u2} (12) & = & \langle \U (1) \cdot \nabla_{m}^{2} \SG (2) \rangle \\
C_{u \Theta}(12) & = & \langle \U (1) \cdot \vec{\Theta} (2)  \rangle.
\eea
We can solve (\ref{EQ:FLUCTCOR1}) and (\ref{EQ:FLUCTCOR2})
to determine the correlations for the fluctuation field $\U$
if we make the additional assumption that $\U$ is also a gaussian
field.
In particular we assume that $\M$ and $\U$ are coupled 
gaussian fields satisfying
%
%
\be
\label{EQ:MIDENTITY} 
\langle m_{i}(1){\cal G}(\M ,\U )\rangle =C_{0}(13)
\langle \frac{\delta}{\delta m_{i}(3)}{\cal G}(\M ,\U )\rangle
+C_{mu}(13)
\langle \frac{\delta}{\delta u_{i}(3)}{\cal G}(\M ,\U )\rangle
\ee
and
\be 
\label{EQ:UIDENTITY}
\langle u_{i}(1){\cal G}(\M ,\U )\rangle =C_{um}(13)
\langle \frac{\delta}{\delta m_{i}(3)}{\cal G}(\M ,\U )\rangle
+C_{uu}(13) 
\langle \frac{\delta}{\delta u_{i}(3)}{\cal G}(\M ,\U )\rangle 
\ee
where ${\cal G}$ is a general function of $\M $ and $\U$, $C_{um}$ is 
defined as
%
\be
\delta_{ij} C_{um} (12) = \langle u_{i}(1) m_{j} (2) \rangle,
\ee
and integrations over ${\bf R}_{3}$ and $t_{3}$ are implied.
Using these identities
all correlation functions depending on $\M$ and $\U$ are
determined in terms of $C_{0}, C_{um}, \mbox{ and } C_{uu}$. 
Thus, we will see that equations (\ref{EQ:FLUCTCOR1}) and (\ref{EQ:FLUCTCOR2}) 
can be expressed in terms of $C_{u0},C_{uu}$ and averages over 
functions of $\M$ alone. This is the first step in determining (\ref{EQ:F1}) 
and (\ref{EQ:F2}). The second step is to evaluate the averages over $\M$, 
which can all be expressed in terms of $C_{0}$, a quantity known from our 
analysis in the last section. The final step is to analyze the equations 
resulting from (\ref{EQ:FLUCTCOR1}) and (\ref{EQ:FLUCTCOR2}) in the 
late-time scaling regime and extract the scaling functions.

We begin by expressing all of the correlation functions involving a power of 
$\U$ which appear in (\ref{EQ:FLUCTCOR1}) and (\ref{EQ:FLUCTCOR2})
in terms of $C_{u0}(12)$, $C_{uu}(12)$ and averages over $\M$.
Using the identity (\ref{EQ:UIDENTITY}) we have 
%
%
\bea
C_{u0}(12) & = & C_{um}(12) M_{1} \label{EQ:CU0CUM} \\
C_{u2}(12) & = & C_{um}(12) M_{3} \label{EQ:CU2CUM} \\
C_{u \Theta}(12) & = & C_{um}(12) \Omega_{1} \\
\langle W_{ij}(1)u_{j}(1)\sigma_{i}(2)\rangle & = &
C_{um}(11)W_{1}(12)+C_{um}(12) W_{2}(12) \\
\langle W_{ij}(1)u_{j}(1)u_{i}(2)\rangle & = &
C_{uu}(12)\Omega_{2} +C_{um}(21) C_{um}(11) \Omega_{3} \label{EQ:WUU} 
\eea
where we define
%
%
\bea
M_{1} & = & \langle \nabla_{m}\cdot\SG (1)\rangle \label{EQ:M1} \\
M_{3} & = & \langle \nabla_{m}\cdot \nabla_{m}^{2} \SG (1)\rangle \\
\Omega_{1} & = & \frac{\A}{L^{2}} M_{1} 
+\sum_{\ell =1}^{\ell_{max}} \omega_{\ell} \langle \nabla_{m}\cdot 
\nabla_{m}^{2} 
\SG (1) [\nabla_{1} \M (1)]^{2(\ell - 1)} \rangle \\
\Omega_{2} & = & \langle W_{ii}(1)\rangle \\
\Omega_{3} & = & \langle \nabla_{m}^{i} \nabla_{m}^{j} W_{ij}(1) \rangle \\
W_{1}(12)  & = & \langle [\nabla_{m}^{j} W_{ij}(1)] \sigma_{i}(2) \rangle \\
W_{2}(12)  & = & \langle W_{ij}(1) \nabla_{m}^{j} \sigma_{i}(2) \rangle.
\label{EQ:W2}
\eea
We see from (\ref{EQ:CU0CUM}) that $C_{um}$ can be eliminated in favor of 
$C_{u0}$ in equations (\ref{EQ:CU2CUM}-\ref{EQ:WUU}). 
In terms of these various auxiliary functions, equations 
(\ref{EQ:FLUCTCOR1}) and (\ref{EQ:FLUCTCOR2}) become
%
%
\bea
\nonumber
\frac{\partial}{\partial t}C_{u0}(12) & = & \nabla^{2}_{1}C_{u0}(12) 
-C_{u0}(11)W_{1}(12)M^{-1}_{1} -C_{u0}(12)W_{2}(12)M^{-1}_{1}
-C_{\Theta 0}(12) \\
& & + \nabla^{2}_{2}C_{u0}(12) -C_{u0}(12)M_{3}M^{-1}_{1}
+ C_{u0 }(12)\Omega_{1}M^{-1}_{1}
\label{EQ:FINALFLUCT1}
\eea
and
\be
\label{EQ:FINALFLUCT2}
\frac{1}{2} \frac{\partial}{\partial t}C_{uu}(12)=
  \nabla^{2}_{1}C_{uu}(12) - \frac{1}{n} C_{uu}(12) \Omega_{2} 
- \frac{1}{n} C_{u0}(12)\bigl(C_{u0}(11)\Omega_{3}M^{-2}_{1} 
+ \Omega_{1}M^{-1}_{1}).
\ee

The next step is to compute the averages over $\M$ (\ref{EQ:M1}-\ref{EQ:W2}) 
in the scaling regime. We begin with the {\it one-point} averages. Except for 
$\Omega_{1}$, which involves spatial gradients, these can all be evaluated 
using 
%
%
\be
\langle A[\M]\rangle =\int\frac{d^{n}x}{(2\pi S_{0})^{n/2}}
e^{-x^{2}/(2S_{0})}A[x].
\ee
It is straightforward to show that
%
%
\be
M_{1} =\psi_{0}\sqrt{\frac{2}{S_{0}(t)}} \frac{\Gamma (\frac{n+1}{2})}
{\Gamma (\frac{n}{2})}
\ee
and
%
%
\be
M_{3} =-\frac{1}{S_{0}(t)}M_{1}.
\ee
Turning to $\Omega_{2}$ we see rather trivially from the form of the
stability matrix given by (\ref{EQ:STABILITY}) that
%
%
\be
\Omega_{2}=q_{0}^{2}+{\cal O}(L^{-2})
\ee
with  a $\ln L$ multiplying the correction term for $n=2$.  This
quantity serves as a {\it mass} for the longitudinal fluctuations
and dominates their determination.  For $\Omega_{3}$ a brief manipulation 
produces the simple result
%
%
\be
\Omega_{3}=(n-1)\frac{q_{0}^{2}}{S_{0} (t)}.
\ee
The last local quantity $\Omega_{1}$, involves averages over spatial 
derivatives.  The key point in handling such quantities is that
$\langle m_{k}(1)\nabla_{i}m_{j}(1)\rangle =0$. One can then rather 
easily derive the recursion relation, valid for $\ell >0$,
%
%
\be
\langle (\nabla \M)^{2\ell}G\rangle = nS_{0}^{(2)}
\biggl[1+\frac{2(\ell -1)}{nd}\biggr]
\langle (\nabla \M)^{2(\ell -1)}G\rangle
\ee
for a general local function $G[\M]$.
With this relation we  can evaluate $\Omega_{1}$, up to the $\omega_{2}$ term:
%
%
\be
\Omega_{1} = M_{1} \left[ \frac{\A}{L^{2}(t)} - \frac{\C}{S_{0}(t)} 
 -\frac{n S_{0}^{(2)} \B}{S_{0} (t)} \right].
\ee

Turning to the {\it two-point} quantities $W_{1}$ and $W_{2}$
it is easy to show, using the symmetry properties of the order-parameter,
that in the scaling regime these reduce to
\be
W_{1}(12)=(n-1)q_{0}^{2}\psi_{0}\langle \frac{\hat{\sigma}(1)
\cdot\hat{\sigma}(2)}{m(1)}\rangle
\ee
and
\be 
W_{2}(12)=q_{0}^{2}\psi_{0}\langle \frac{1}{m(2)} [
1- (\hat{\sigma}(1) 
\cdot\hat{\sigma}(2))^{2} ]\rangle. 
\ee
These are new averages to be evaluated.  Hereafter, we shall work exclusively
with $n=2$. In this case $W_{1}(12)$ and $W_{2}(12)$ can be evaluated as
\be
W_{1}(12)=q_{0}^{2}\psi_{0}\sqrt{\frac{\pi}{2S_{0}}}
\frac{1}{f} (1-\sqrt{1-f^{2}})
\ee
while
\be 
W_{2}(12)=q_{0}^{2}\psi_{0}\sqrt{\frac{\pi}{2S_{0}}} 
\frac{\sqrt{1-f^{2}}}{1+\sqrt{1-f^{2}}}. 
\ee

We are now in a position to evaluate $C_{u0}$ and $C_{uu}$ in
the scaling regime, and to relate them to $\F$ through $C_{\Theta 0}$. 
From the 
definition (\ref{EQ:THETADEF}) of $\vec{\Theta}$ we see that the scaling 
ansatz for $C_{\Theta 0}$ should have the form
%
%
\be
\label{EQ:CT0}
C_{\Theta 0}=\frac{\psi_{0}^{2}}{L^{2}}F_{\Theta}(x)
\ee
where, after explicit evaluation,
%
%
\be
F_{\Theta} = \A \F - \frac{\pi}{2 \mu S_{0}^{(2)}} (\C +2 S_{0}^{(2)} \B)
f \partial_{f} \F + \B [\nabla_{x} f]^{2} [f \partial_{f} \F + f^{2} 
\partial_{f}^{2} \F ].
\ee
Looking at the determining equations (\ref{EQ:FINALFLUCT1}) and 
(\ref{EQ:FINALFLUCT2}) we easily see that, as a consequence of (\ref{EQ:CT0}),
we must take $ u \sim L^{-2}$ to leading order. We therefore write
\be
C_{u0}(12)=\frac{\psi_{0}^{2}}{L^{2}}F_{u}(x)
\ee
and
\be
\label{EQ:FUUSCALE} 
C_{uu}(12)=\frac{\psi_{0}^{2}}{L^{4}}F_{uu}(x).
\ee
With these ansatze (\ref{EQ:FINALFLUCT1})  can be written as
\be
F_{u}(0)\frac{1}{f} (1-\sqrt{1-f^{2}})
+F_{u}(x)\frac{\sqrt{1-f^{2}}}{1+\sqrt{1-f^{2}}}=-\frac{1}{q_{0}^{2}}
F_{\Theta }(x)
\ee
to leading order, while (\ref{EQ:FINALFLUCT2}) becomes
\be
F_{uu}(x)+\frac{2}{\pi}F_{u}(0)F_{u}(x)= -\frac{1}{q_{0}^{2}} \biggl[\A 
-\frac{\pi}{2 \mu S_{0}^{(2)}} (\C + 2 S_{0}^{(2)} \B )\biggr] F_{u}(x).
\ee
We see that the quantity $F_{u}(0)$ enters into these equations.  If
this quantity is to be finite then we see that $F_{\Theta }(x)$ can
not blow up as $x \rightarrow 0$.  Since $f \partial_{f} \F$ does blow
up as $x \rightarrow 0$ we must choose
%
%
\be
\C + 2 S_{0}^{(2)} \B = 0 
\ee
which fixes $\C$ in terms of $\B$ and tells us, using (\ref{EQ:S02DEF}),
that $S_{0}^{(2)}=1$ for $n=2$, even in the presence of these perturbations.  
We then find that
\bea
\nonumber
F_{u}(0) & = & \frac{2 \B f_{2} - \A}{q_{0}^{2}} \\
         & = & - \frac{2 f_{2} (1 + d K_{2})}{q_{0}^{2}},
\label{EQ:FU(0)}         
\eea
where (\ref{EQ:C0C2K2}) has been used.
We then have the final results 
\be
\label{EQ:FU}
F_{u}(x)=- \frac{\gamma}{q_{0}^{2}}
\left[ \A [ (1+\sqrt{1-f^{2}})\F-f ] + 2 \B f_{2} f 
+\B [1 + \sqrt{1 - f^{2}}][\nabla_{x}f(x)]^{2}[f \partial_{f} \F + f^{2} 
\partial_{f}^{2} \F] \right]
\ee
and
\be 
%
%
%
%
\label{EQ:FUU}
F_{uu}(x) = - \frac{1}{q_{0}^{2}} \left[ \A + \frac{2 q_{0}^{2}}{\pi} 
F_{u}(0) \right] F_{u}(x). 
\ee
Inspection of equations (\ref{EQ:FU(0)}), (\ref{EQ:FU}) and (\ref{EQ:FUU}) 
show that in the original theory \cite{LIU92a}  $F_{u}(x) = F_{uu}(x) = 0$,
as expected.
From the definitions (\ref{EQ:F2}) and (\ref{EQ:FUUSCALE}) we must have 
$F_{uu}(0) \geq 0$. In the theory with only $\A \neq 0$ we have
%
\be
F_{uu}(0) = \frac{\A^{2}}{q_{0}^{4}} \left[1 - \frac{2}{\pi} \right] 
\ee
which is positive. However, if $\A=0$ (\ref{EQ:FUU}) implies 
%
%
\be
F_{uu}(0) = - \frac{2}{\pi} [ F_{u}(0) ]^{2}
\ee
which is negative. 
Thus within the $\ell_{max} = 2$ approximation it is necessary to have 
$\A \neq 0$ in order to have a physical theory.
For more general ${\vec \Theta }$, one must look to the numerical solution of 
(\ref{EQ:SCALE}) to answer the question of the sign of $F_{uu}(0)$.

Equations (\ref{EQ:FU}) and (\ref{EQ:FUU}) explicitly show how 
correlations in the $\U$ field are slaved to those of the order-parameter. 
The universality in (\ref{EQ:FU}) and (\ref{EQ:FUU}) is evident, up to the
non-universal overall factor of $1/q_{0}^{4}$, which characterizes the 
flatness of the equilibrium minimum in the potential and sets the scale of 
the fluctuations.

\section{Numerical analysis of the non-linear eigenvalue problem}

The eigenvalue problem posed by (\ref{EQ:SCALE}), subject to the boundary 
conditions at small- and large-$x$ outlined above, has to be solved 
numerically.
A fourth-order Runge-Kutta integrator is used to integrate (\ref{EQ:SCALE})
with initial conditions given by an analytic small-$x$ expansion to 
$x = 0.0001$. The eigenvalue $\mu$ is adjusted until the solution matches 
onto the gaussian decay  (\ref{EQ:LARGEX}) at the largest distances. This is 
now a standard procedure and is essentially the same as that used in 
\cite{LIU92a}. We examine the $O(2)$ model in two and three spatial dimensions.

In the original theory \cite{LIU92a} $\vec{\Theta} = 0$ and the selection
of the eigenvalue depended on only the conditions outlined 
above. In that theory $K_{2} = -1/d$, which lead to an unsatisfactory 
non-analyticity in the small-$x$ behaviour of $f(x)$ and ultimately to an 
unphysical divergence in $\tilde{g}(x)$ at small $x$.  In the present theory 
we can choose $\vec{\Theta}$ so as to eliminate the leading 
non-analyticities and remove the unphysical divergence. Each choice 
represents a separate eigenvalue problem. The simplest choice is to keep only
the first term in in $\vec{\Theta}$ by setting $\A = 2 f_{2}$ and 
 $\omega_{\ell} = 0 \mbox{ for } \ell > 0$. For $d=2$ the solution to this 
eigenvalue problem has $\A = -1.4620...$, while for $d=3$ one has 
$\A = -1.3450...$. Another choice for 
$\vec{\Theta}$ is to eliminate the first term, but 
keep the next two by setting  $\A = 0, \mbox{ } \C = 2, \mbox{ } \B = -1 
\mbox{ and } 
\omega_{\ell} = 0 \mbox{ for } \ell > 2$. As mentioned in the previous 
section, this choice has the unfortunate consequence of rendering $F_{uu}(0)$
negative. Both these theories have $K_{2} =0$ and $K_{4} \neq 0$. 
If we require $f(x)$ to be analytic up to ${\cal O}(x^{4})$ we must choose 
all of $\A, \mbox{ } \C$ and $\B$ to be non-zero, following the prescription 
outlined in Section IV to ensure that $K_{2}=K_{4}=0$.  The solution to 
this eigenvalue problem has $\A = -2.6004...$, $\B = 0.41914...$ 
for $d=2$ and $\A = -2.3438...$, $\B = 0.45675...$ for $d=3$.

Table \ref{TBL:MU} contains the eigenvalues $\mu$ obtained 
from these theories. The autocorrelation exponents are shown in Table 
\ref{TBL:LAMBDA}. The scaling forms for the order-parameter correlation 
functions $\F$ from the various theories are compared in Figure 
\ref{FIG:CORRELATION} for the $O(2)$ model in two dimensions. 
The three dimensional results for $\F$ are similar, and are not shown. 
The defect-defect correlation function 
$\tilde{g}(x)$ (\ref{EQ:DEFECTDEFECTCOR}) is calculated from our 
numerically determined form for $f(x)$. The results for the correlations
between vortices $(n=d=2)$ obtained from the various
theories are compared in Figure \ref{FIG:DEFECT}. Simulation results 
\cite{MONDELLO90}, which are scaled to give the best fit to the original 
theory \cite{LIU92a} at large $x$, are shown for comparison. 
Finally, the scaling function  $F_{uu}$ (\ref{EQ:FUU}) 
is computed using our knowledge of $\F$ and $f$. The results for the 
theory with $\A \neq 0$ and $\B \neq 0$ are shown in Figure \ref{FIG:FUU} 
for two and three spatial dimensions. For the theory with just $\A \neq 0$
the behaviour of $F_{uu}$ is similar to that shown.

\section{Discussion}

It is noteworthy that 
the scaling solution for we find here for $\F$ retains the same qualitative 
features seen
in \cite{LIU92a}.  The differences between the $\F$'s can 
be attributed mainly to a rescaling of the length scale. Since there is always
some arbitrariness in the choice of length scale when comparing simulation 
data to theory, we expect that our new results for $\F$ will fit the 
simulation data well after rescaling. To our knowledge, simulation 
results for the autocorrelation exponent $\lambda$ of the $O(2)$ 
model exist only for two spatial dimensions \cite{LEE95}, where it is found
that $\lambda = 1.171$. We see from Table \ref{TBL:LAMBDA} that the 
original theory \cite{LIU92a} is already in excellent agreement with the 
simulations on this point. It is not surprising then that the modified theory
makes worse predictions for $\lambda$ than the original theory. The 
introduction of $\vec{\Theta}$ was not expected to be in any sense a small 
perturbation. A trend which {\em is} counter to our expectations is that the
discrepancy between the value for $\lambda$ from simulations and the value 
obtained in the theory seems to increase as
one includes more terms in $\vec{\Theta}$. It may still turn out that the 
inclusion of higher order terms in $\ell$ in $\vec{\Theta}$  does lead to 
improvement. This appears to be a 
straightforward but tedious calculation. It is also interesting to note that,
for the theory with $\A = 0$ and $\B \neq 0$, the prediction for $\lambda$ 
violates a 
proposed \cite{FISHER88} lower bound $\lambda > d/2$. Despite these problems
with the quantitative values obtained for $\lambda$, there is qualitative 
improvement in 
$\tilde{g}(x)$ since the small-$x$ divergence for $n=d=2$ seen in the original 
theory is removed. The value of $\tilde{g}(0)$ in all the modified theories 
is too low when compared with simulations and this point suggests that some 
fine-tuning of the theory is 
necessary. Finally, our results for $F_{uu}$ obey the necessary condition
$F_{uu}(0) > 0$ for two of our choices of $\vec{\Theta}$. We also see that the
strength of the fluctuations, characterized by $F_{uu}(0)$, increases in 
lower dimensions, as one might expect.

We see that the inclusion of fluctuations allows us to render the correlation
function 
$f$ of the auxiliary field analytic and we have constructed such a solution 
up to ${\cal O}(x^{4})$. This, in turn, cures the short-distance divergence in
$\tilde{g}$. The theory, at the gaussian level, appears to be in better 
qualitative shape given our development here, however it is at the expense 
of proper quantitative agreement for the non-equilibrium exponent $\lambda$.
There is evidence \cite{WICKHAM} that the inclusion of post-gaussian 
corrections lowers the value of $\lambda$. Thus we hope that the tendency 
for the fluctuations to increase $\lambda$ will be balanced by the 
introduction of post-gaussian terms, resulting in a value for 
$\lambda$ in reasonable agreement with simulations. We also 
hope that post-gaussian corrections will reduce the magnitude 
of $\tilde{g}(0)$. It appears that the procedure we introduce here leads to 
a qualitatively more consistent theory. However it is also clear that it is 
unlikely that one can have such a theory and quantitative estimates for 
exponents within the gaussian approximation. One should proceed to look at 
post-gaussian theories.
%
%
\acknowledgements
This work was supported in part by the MRSEC Program of the National Science
Foundation under Award Number DMR-9400379. 
R.A.W. gratefully acknowledges support from the NSERC of Canada. 
%
%

%
%
\begin{table}
\caption{Values for the eigenvalue $\mu$ from the various theories 
(in all theories $\C = - 2 \mbox{ } \B$).}
\label{TBL:MU}
\begin{tabular}{ccccc}
        & $ \A = \B = 0$   &  $\A \neq 0$, $\B = 0$ & $\A = 0$, $\B = -1$
  & $\A,\mbox{ }\B \neq 0$ \\
\tableline
%
%
%
%
$d = 2$  & 0.94858 & 0.53721 & 0.76033 & 0.42863 \\
$d = 3$  & 0.56837 & 0.38931 & 0.51434 & 0.32544 \\
\end{tabular}
\end{table} 
%
%
\begin{table}
\caption{Values for the autocorrelation exponent  $\lambda$ 
from the various theories (in all theories $\C = - 2 \mbox{ }\B$).}
\label{TBL:LAMBDA}
%
\vspace{.5in}
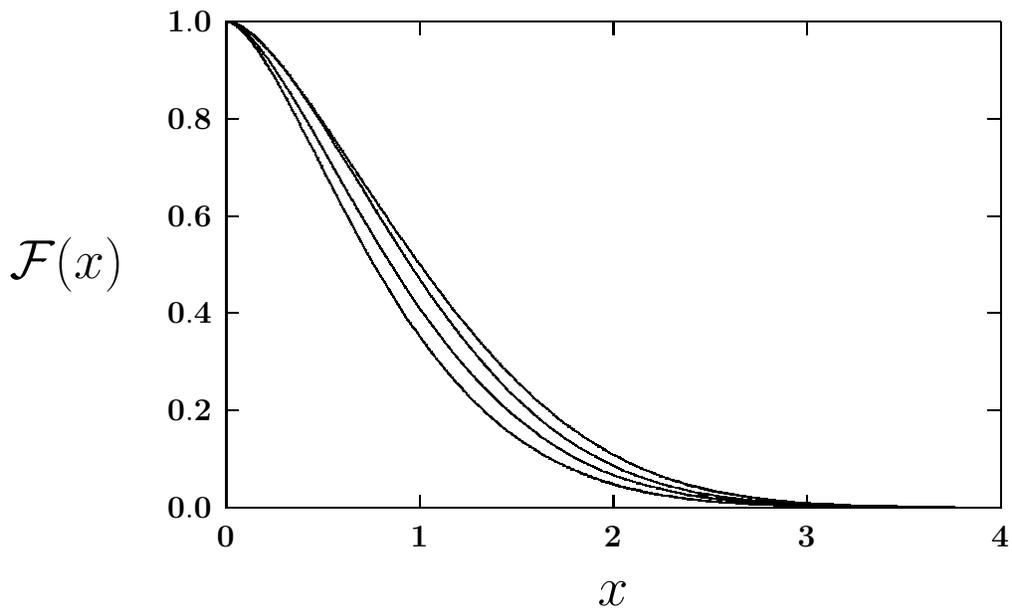
\caption{Scaling function $\F(x)$ for the order-parameter correlations in 
two dimensions. From bottom to top, at $x = 1$, the curves correspond to: the 
theory with $\A \neq 0$ and $\B \neq 0$; the theory with only $\A \neq 0$; 
the unmodified theory \protect\cite{LIU92a}; the theory with $\A = 0$ and $\B
 \neq 0$. In all theories $\C= -2 \mbox{ } \B$.}
\label{FIG:CORRELATION}
\end{figure}
\pagebreak
\begin{figure}
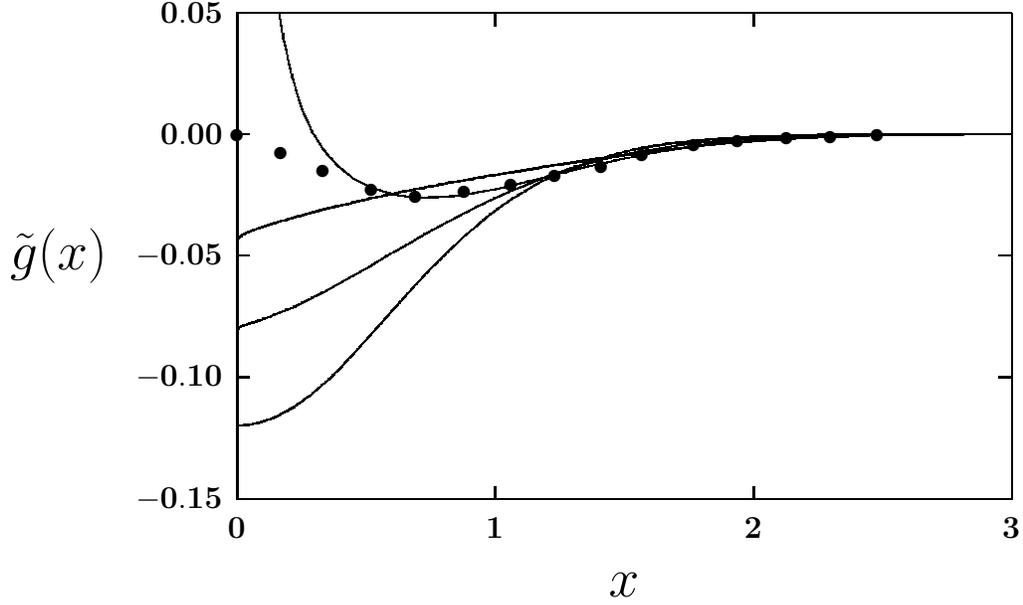

\setlength{\unitlength}{0.240900pt}
\ifx\plotpoint\undefined\newsavebox{\plotpoint}\fi
\sbox{\plotpoint}{\rule[-0.200pt]{0.400pt}{0.400pt}}%

\vspace{.5in}
\caption{Scaling function $\tilde{g}(x)$ for the defect-defect correlations
in two dimensions. From bottom to top at $x =0$ the solid curves 
correspond to: the theory with $\A \neq 0$ and $\B \neq 0$; the theory with 
only $\A \neq 0$;  the theory with $\A = 0$ and $\B \neq 0$, the unmodified 
theory \protect\cite{LIU92a} (diverging). In all theories $\C= -2 \mbox{ } 
\B$. The dots represent the simulation results \protect\cite{MONDELLO90} for 
the two-dimensional $O(2)$ model.}
\label{FIG:DEFECT}
\end{figure}
\pagebreak
\begin{figure}
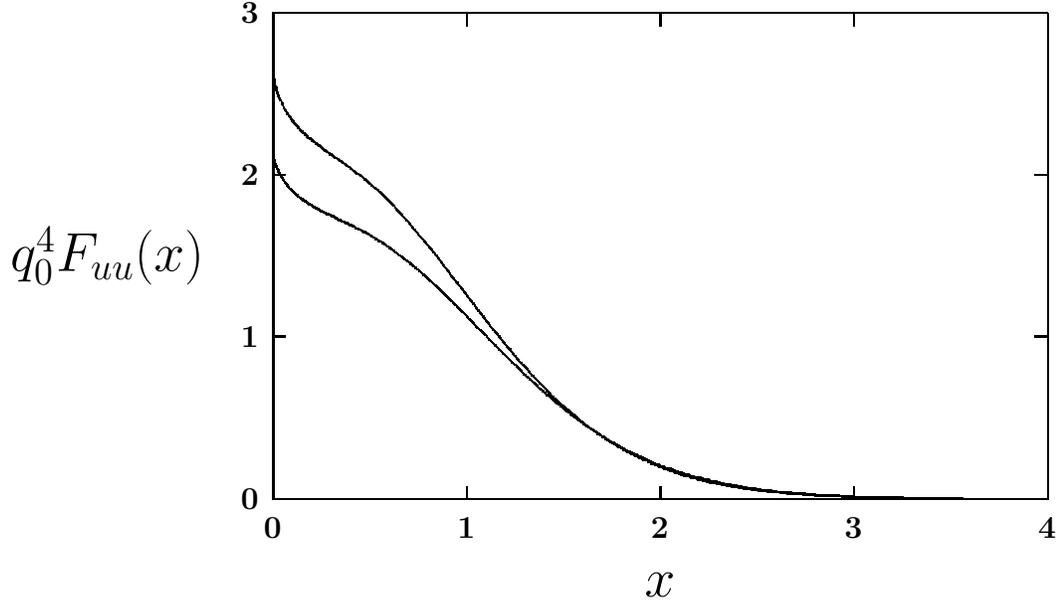

\setlength{\unitlength}{0.240900pt}
\ifx\plotpoint\undefined\newsavebox{\plotpoint}\fi
\sbox{\plotpoint}{\rule[-0.200pt]{0.400pt}{0.400pt}}%

\vspace{.5in}
\caption{Scaling function for the fluctuation correlations for the theory 
with $\A \neq 0, \mbox{ } \B \neq 0$ and $\C =  -2 \mbox{ } \B$. 
At $x=0$ the lower curve is the result for three dimensions and the upper 
curve is the result for two dimensions.}
\label{FIG:FUU}
\end{figure}

\begin{thebibliography}{99}
\bibitem{CONFUSION} The gaussian form for $f$ is completely unrelated to the 
gaussian statistics for $\M$.
\bibitem{POTENTIAL} Typically, $V$ is taken to have the form:
                    $V = \frac{1}{4} (\psi_{0}^{2} - \psi^{2})^{2}$.
\bibitem{MASS} There is a mass term in the equation for $\U_{T}$, but it is 
               ${\cal O} (1/L^{2})$.
\end{thebibliography}
\end{document}